# Heterogeneous Multi Core Processors for Improving the Efficiency of Market Basket Analysis Algorithm in Data Mining

Aashiha Priyadarshni.L

*B.Tech Information Technology, College of Engineering Guindy*
*Anna University, Chennai-600 025, Tamil Nadu, India*

*Abstract---* **Heterogeneous multi core processors can offer diverse computing capabilities. The efficiency of Market Basket Analysis Algorithm can be improved with heterogeneous multi core processors. Market basket analysis algorithm utilises apriori algorithm and is one of the popular data mining algorithms which can utilise Map/Reduce framework to perform analysis. The algorithm generates association rules based on transactional data and Map/Reduce motivates to redesign and convert the existing sequential algorithms for efficiency. Hadoop is the parallel programming platform built on Hadoop Distributed File Systems(HDFS) for Map/Reduce computation that process data as (key, value) pairs. In Hadoop map/reduce, the sequential jobs are parallelised and the Job Tracker assigns parallel tasks to the Task Tracker. Based on single threaded or multithreaded parallel tasks in the task tracker, execution is carried out in the appropriate cores. For this, a new scheduler called MB Scheduler can be developed. Switching between the cores can be made static or dynamic. The use of heterogeneous multi core processors optimizes processing capabilities and power requirements for a processor and improves the performance of the system.**

## I. INTRODUCTION

Heterogeneous multi core processors provide a new trend for various computing capabilities. As the processor speed and performance increases, the main challenges found today are processor power consumption and heat dissipation. Heterogeneous multi core processors can sufficiently reduce processor power consumption and can considerably increase the speed of execution and performance. The multi core processors reduce the frequency of the processors and thus reduce the temperature of the system. In these processors the instructions run simultaneously on individual cores and the amount of parallelism is increased [1]. When the application is executed, the system software can dynamically choose the most appropriate core to satisfy the performance and power requirements. Thus the use of heterogeneous multi core processors optimizes processing capabilities and power requirements for a processor.

Market Basket Analysis algorithm helps us to analyse customer purchasing patterns by extracting associations or co-occurrences from the transactional database in a store. This algorithm involves generation of association rules based on the information obtained from the analysis and helps in sales, marketing, service and operation strategies. The Market Basket Analysis algorithm uses Apriori Algorithm for analysing and mining association rules from the transactional database[2].

Hadoop is the parallel programming platform built on HDFS[Hadoop Distributed File Systems] for performing Map/Reduce computation that process data as (key, value) pairs. HBase is one of the NO-SQL Databases which runs on HDFS with Hadoop and can store and process Big Data. Market Basket Analysis Algorithm runs on Hadoop Map/Reduce and the transactional data can be stored in HDFS. When large amount of transactional data need to be stored and processed, HBase can be used [3].

When we replace homogenous cores with heterogeneous core processors, we need to take into account of various performance parameters and factors like RAM Optimisation, Scheduling, Memory Consumption, Cache utilisation, Core switching and power consumption and so on. In this paper performance of factors like Scheduling, Core switching and power consumption can be optimised.

## II. PROBLEM STATEMENT

This paper aims to execute Market Basket analysis algorithm under the heterogeneous multi core environment using Hadoop Map Reduce framework. The algorithm must be split into tasks and executed in Map Reduce. Then each task must be assigned appropriate core for execution. To enable this functionality, we assume a new scheduler called MB Scheduler. The MB Scheduler can be used to optimise execution in heterogeneous multi core. Thus the efficiency of Market Basket Analysis Algorithm can be sufficiently improved.

## III. MARKET BASKET ANALYSIS USING MAP REDUCE FRAMEWORK

Market Basket Analysis (Aprioi Algorithm) is transformed to a distributed structure and applied to the map-reduce framework in hadoop and executed. In hadoop framework, the master node (Job tracker) assigns task to the worker nodes (task tracker). The worker nodes perform the task and return the results to the master node [4].

The apriori algorithm is executed in the hadoop environment. The input (transactional data) may be stored in HDFS or HBase depending upon the size. Map/Reduce involves three steps in computing the apriori algorithm. Input Data collected from the transactional database are stored in HDFS or HBase. Generally in the Map phase, master node maps the input dataset into N partitions and sends them to the worker nodes [N must equal to the number of worker nodes]. In the worker nodes, Key-Value pair is generated for each item. The Reduce phase collects the intermediate key-value pairs emitted in map phase and sends them to the master node.

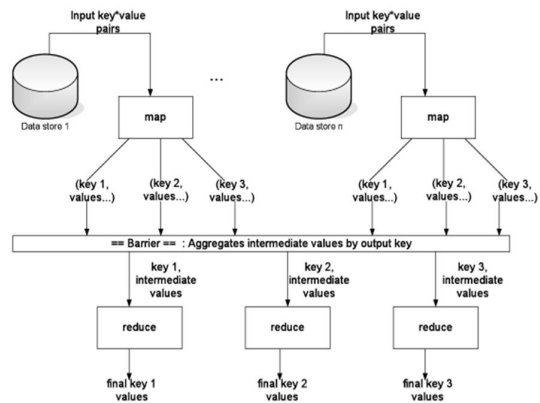

Fig.1 - A MapReduce computation [8].

Market Basket Analysis in hadoop involves three steps. Each step consists of a map phase and a reduce phase. In the first step, the frequency of the items is collected. In the Map phase, master node generates the Key-Value pair for each item in form of <item, count>. The Reduce phase collects the item-count pairs emitted in map phase and generates the file of count per item. In the second step, the candidate generation takes place. The Map phase prunes the items based on minimum support value and generates candidates. The Reduce phase groups candidate item set and collects their count. Finally, in the third step candidate item set is given as input. The Mapper prunes and generates association rules based on minimum confidence values. The Reduce phase collects all association rules and returns the result to the master node [5]. Thus market basket analysis can be performed through map/reduce framework [8].

## IV. HETEROGENEOUS MULTI CORE FRAMEWORK

The heterogeneous multi core framework consists of series of cores with different processing capabilities. When the application needs to be executed it chooses the most appropriate core. The master node parallelizes and assigns tasks to be executed to the worker nodes. The tasks can be executed in the heterogeneous multi core environment. When the applications (tasks) get executed, they can choose the appropriate core among various cores for execution. The selection of the cores may be based on various factors like performance, speed, power consumption, RAM optimisation, scheduling and so on [6,7].

The parallelized tasks can be either single threaded or multi threaded. In case of single threaded application, an application can choose only one core at a time. It chooses the best core depending upon the tasks involved. The other cores can be switched off to reduce power requirements. In case of multithreaded applications, various threads of same application can choose different appropriate cores dynamically or statically and perform their execution simultaneously.

In Market Basket Analysis algorithm, the map/reduce tasks allocated to the task tracker (worker nodes) are executed under the heterogeneous multi core environment. The mapper submits tasks to the processors and executes them. The Reducer collects the result from the processors and combines them to form the output [7]. The task assigned to the various processors can be maintained through a scheduler.

## V. METHODOLOGY

We assume a system with heterogeneous multi core environment which involves multiple cores which are heterogeneous. Since Heterogeneous Environment is not available commercially, we have considered a Hadoop Cluster with different cores which can serve as a heterogeneous multi core system. Consider a system consisting of four cores with different processing powers like 80MB, 120MB, 200MB and 400MB. A scheduler can be used to instruct the mapper to assign tasks to the four cores and instruct the reducer to retrieve outputs from them after execution. For this we assume a scheduler called MB Scheduler.

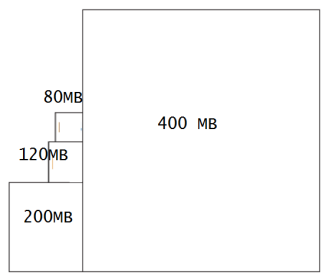

*Fig.2* Heterogeneous Multi Core System

The MB Scheduler can be used to assign the tasks to the processor and collect the output from them after execution. The transactional data is stored in either HDFS or HBase depending upon its size. The Job of extracting associations or co-relations from the transactional data is submitted to the system. When the job is assigned to the Job Tracker in the system, the job tracker analyses the job and allocates the task (job is split into tasks) to the task tracker. Using MB Scheduler, the task tracker analyses the tasks and mapper allocates them to the appropriate core for execution. The market basket algorithm thus is executed in a map reduce framework. The mapper and reducer are used to generate <key, value> pairs. The three steps are involved in execution.

Step 1: Frequency of items is collected. Mapper generates Key-Value pair for each item. Reducer generates count per item.
Step 2: Candidate itemset is generated. Mapper prunes items and generates candidates based on minimum support value. Reducer groups candidate itemset and collects their count.
Step 3: Association rules are generated. Mapper prunes candidate itemset and generates association rules based on minimum confidence values. The Reduce phase collects all association rules and returns the result to the task tracker.

The functions of the MB Scheduler can be the following:
1. Collect the tasks submitted to the task tracker.

2. Analyses the tasks whether they are single threaded or multi threaded.

3. In case of single threaded, it analyses processing power required and the mapper assigns the task to the most optimised core. The other cores are completely switched off to reduce the power requirements. A single threaded task can be switched between the cores when required. During that time, before switching into next core, data in the current is stored in the cache. The current core is completely switched off. The new core collects data from the cache and begins execution.

4. In case of multi threaded task, the task is broken into many threads and the threads are run parallel on four cores. After the execution is over, the MB Scheduler collects all the sub tasks from all the cores and combines them to get the output. Here all the processor cores are run simultaneously and the shared data can be stored in a cache.

5. After execution, the reducer collects output from the processors and combines them and returns the output to the task tracker.

The main aim is to run the algorithm on heterogeneous multi core environment so that efficiency of the algorithm can be improved. But the cost of execution should not exceed the benefits of using heterogeneous multi core.

## VI. CHALLENGES IN USING HETEROGENEOUS MULTI CORE

While using heterogeneous multi core processors various factors like power consumption, scheduling, core switching, RAM optimisation has to be taken care of. To decrease the power consumption we can switch off the unused cores completely and perform the execution of the application. The cost for Core switching should not exceed the cost incurred in using heterogeneous multi core processors. Core switching can be done either statically or dynamically. If the order of execution of the cores is known already, then the order can be maintained in a queue and switching between the cores can be made static. Dynamic switching between the cores can be performed using MB Scheduler. The MB Scheduler can be used to keep track of the sub tasks assigned to various cores and can collect the results and combine them after execution based on order.

Scheduling the tasks to various cores must be optimised. The processing power requirements of the tasks must be carefully analysed by the scheduler. The scheduler analyses the amount of data need to be processed and the various algorithms that need to be executed and the time of execution and calculates the processing power requirements. In order to increase the performance, the MB Scheduler must choose the most optimised and appropriate core for execution.

## VII. CONCLUSION

Replacing homogeneous cores with heterogeneous cores can considerably reduce power consumption and can considerably improve the performance and efficiency of the algorithm. The various results and association rules obtained after execution are used to find various purchasing patterns involved in the transactions. Thus efficiency of market basket algorithm can be sufficiently improved by using heterogeneous multi core processors.

FUTURE WORK

Though Heterogeneous multi core architecture imposes few challenges, it improves the performance capabilities for executing market basket analysis algorithm. In the future work, we focus on decreasing the challenges like RAM optimisation, decreasing the processing time and so on. Today, power crisis is one of the major problems faced by the world. Using heterogeneous multi core we can considerably reduce the power consumption and increase the performance. Our future work will focus on decreasing the cost and challenges incurred in using heterogeneous multi core for market basket analysis and aims to improve the performance to a wider extent.

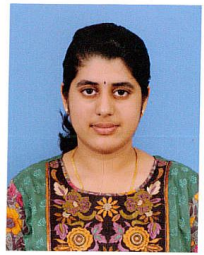

**Author: Aashiha Priyadarshni** is currently B.Tech student in Information Technology from College of Engineering Guindy Anna University. Her research interested areas are in Data Mining, Big Data and Cloud Computing.